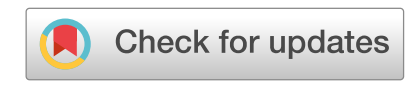

OPEN

# Fabrication of Al/$AlO_x$/Al junctions with high uniformity and stability on sapphire substrates

Yuzhen Zheng[1,2,4], Shuming Li[3,4], Zengqian Ding[3], Kanglin Xiong[2,3]✉, Jiagui Feng[2,3]✉ & Hui Yang[1,2,3]✉

**Tantalum and aluminum on sapphire are widely used platforms for qubits of long coherent time. As quantum chips scale up, the number of Josephson junctions on sapphire increases. Thus, both the uniformity and stability of the junctions are crucial to quantum devices, such as scalable superconducting quantum computer circuit, and quantum-limited amplifiers. By optimizing the fabrication process, especially, the conductive layer during the electron beam lithography process, Al/$AlO_x$/Al junctions of sizes ranging from 0.0169 to 0.04 μm$^2$ on sapphire substrates were prepared. The relative standard deviation of room temperature resistances ($R_N$)−$\sigma_{R_N}/\langle R_N \rangle$ of these junctions is better than 1.7% on 15 mm × 15 mm chips, and better than 2.66% on 2 inch wafers, which is the highest uniformity on sapphire substrates has been reported. The junctions are robust and stable in resistances as temperature changes. The resistances increase by the ratio of 9.73% relative to $R_N$ as the temperature ramp down to 4 K, and restore their initial values in the reverse process as the temperature ramps back to room temperature. After being stored in a nitrogen cabinet for 100 days, the resistance of the junctions changed by1.16% on average. The demonstration of uniform and stable Josephson junctions in large area paves the way for the fabrication of superconducting chip of hundreds of qubits on sapphire substrates.**

While the second quantum revolution is unfolding, it is very urgent to exploit the wide applications of various of superconducting quantum devices. Josephson junction is a device consisting of two superconductors separated by a thin insulator with a few nanometers[1]. The tunnel junction has the characteristics of low loss and strong nonlinearity, and plays essential roles in quantum devices, including superconducting qubit, single microwave photon detectors, and quantum-limited amplifiers[2–6]. Since there is a direct relationship between the frequency of the qubit and the $R_N$[7], for multi-qubit chips, the variations of $R_N$ of the Josephson junction may lead to frequency collisions between qubits. Additionally, the non-uniformity of critical current can lead to unwanted reflections in the Josephson traveling wave parametric amplifier and reduce device performance[6]. Preparing Josephson junctions on wafer scale with high uniformity and stability with common facilities is very important.

It is challenging to fabricate wafer-scale highly uniform Josephson junctions, especially on sapphire. Researchers have made a lot of efforts to improve the uniformity of Al/$AlO_x$/Al junctions on high-resistivity silicon substrate. By optimizing the fabrication process, it is reported that, 3.5% resistance variation for 0.042 μm$^2$ Al/$AlO_x$/Al junctions on a 49 cm$^2$ chip[8]; 3.7% resistance variation for Al/$AlO_x$/Al junctions on a wafer that contains forty 0.5 × 0.5 cm$^2$ chips[9]; and 3.9% critical current variation for Al/$AlO_x$/Al junctions on a 20 × 20 mm$^2$ chip[10]. To further adjust of the resistance, laser annealing was developed[7,11]. The methods used on silicon may not work for sapphire. Sapphire is a commonly used substrate for superconducting quantum circuits due to its very low microwave loss, and compatible with growth of low loss materials like tantalum. The longest coherence time for a superconducting qubit has been reported on sapphire[12]. However, it is not only difficult to achieve uniform junction patterns using low-energy electron beam exposure (due to charging effect), but also hard to improve uniform junction resistance using laser annealing (due to transparency to light). Therefore, exploring the fabrication process for Al/$AlO_x$/Al junctions with high uniformity on a large scale on sapphire is critical for developing high-quality superconducting quantum processors[13,14].

[1]School of Nano-Tech and Nano-Bionics, University of Science and Technology of China, Hefei 230026, People's Republic of China. [2]Suzhou Institute of Nano-Tech and Nano-Bionics, Chinese Academy of Sciences, Suzhou 215123, People's Republic of China. [3]Gusu Laboratory of Materials, Suzhou 215123, People's Republic of China. [4]These authors contributed equally: Yuzhen Zheng and Shuming Li. ✉email: klxiong2008@sinano.ac.cn; jgfeng2017@sinano.ac.cn; hyang2006@sinano.ac.cn

In this work, the preparation process of Al/$AlO_x$/Al junction on sapphire substrate was systematically explored. Then the uniformity and stability of their junction resistances were studied. Larger accelerating voltage of electron beam exposure has smaller forward scattering[15], which means that the electrons are more likely to travel straight through the electron beam resist without being deflected. However, the 100 kV electron beam lithography (EBL) is not available to many labs. Using the electron beam exposure with the maximum accelerating voltage of 50 kV, by optimizing the fabrication process, especially, the conductive layer during the electron beam lithography process, Al/$AlO_x$/Al junctions were fabricated in the sizes of ranging from 0.0169 to 0.04 μm$^2$. These junction resistances show high uniformity with $\sigma_{R_N}/<R_N>$ better than 1.7% on 15 mm × 15 mm chips, and $\sigma_{R_N}/<R_N>$ better than 2.66% on 2 inch wafers, which is the highest uniformity on sapphire substrates has been reported. Furthermore, we find that these junctions exhibit robust stability in resistances, whose resistance increase by 9.73% relative to $R_N$ as the temperature decreases from room temperature (300 K) to 4 K, and almost return to their initial values in a reversible process when the temperature rises back. This is consistent with the existing reports[16]. After being stored in a nitrogen cabinet for 100 days, the resistances of these junctions changed very little. This paves the way for the preparation of nearly 100-qubit superconducting circuit with long qubit coherence time based on sapphire substrates.

## Fabrication

Al/$AlO_x$/Al junctions with the characteristic linewidth ranging from 130 to 200 nm have been fabricated on 2-inch c-plane sapphire substrates. The area of the traditional Dolan style junction[17] is dependent on the thickness of the resist and the deposition angle of the bottom and top electrodes, which can affect the uniformity of the junctions. The bridgeless 'Manhattan Style' junctions[18,19] were used in this report. Additionally, in order to avoid introducing two level systems (TLS) and other unstable factors in the 'parasitic' junctions that can cause parameter fluctuations[20], the technique called 'Patch integrated cross-type (PITC)' was used[9]. This technique enables the fabrication of both the junction and the patch in a single lithography step by evaporating from three optional angles. To prepare submicron Al/$AlO_x$/Al junctions, layouts were generated and exposed by using a 50 kV Electron Beam Pattern Generator. After the pattern transfer is completed by photolithography, junctions are deposited in Plassys MEB550SL3 with base pressure of $3\times10^{-8}$ mbar.

The bi-layer electron beam resist used 500 nm MMA EL9 as the bottom resist and 300 nm PMMA A4 as the top resist. In order to spin MMA more uniformly, the small hole in the top of the spin coater was covered when spinning MMA[8]. In the process of electron beam exposure, poor conductivity of the sapphire substrate can lead to charge accumulation. The accumulated charges induce electric fields on the surface of the sample, causing deflection of primary and secondary electrons, which can reduce the pattern resolution and positioning precision[21], resulting in poor uniformity of junctions. To reduce the charging effect, covering the photoresist with charge dissipaters[22] is a good solution, but it may cause contamination. To avoid subsequent contamination, we chose Al, which is easily removable. The Al layer is deposited at room temperature with a deposition rate of 1.2 nm/s by sputtering. However, a thick conductive layer increases electron scattering volume, resulting in decreased resolution. Therefore, the thickness of the conductive layer should be as thin as possible. Insufficient conductivity of a thin conductive layer still causes significant charging effects, as shown in Fig. 1c (top), leading to obvious distortion and poor edges of the Josephson junctions produced. When a conductive layer with 20 nm was used, we obtained Josephson junctions with steep edges (Fig. 1c (bottom)). In this experiment, both MMA and PMMA are exposed by 50 kV electron beam, and the optimal exposure doses were 200 and 1100 μC/cm$^2$ respectively. To remove the Al conductive layer after the exposure, a two-step method was developed. Firstly, a diluent of TMAH with minimal attack on the electron beam resist was used to etch most of the Al layer, and the remaining was removed with deionized water, then the final patterns were defined. The sample was developed at room temperature with IPA: MIBK = 3: 1. Oxygen plasma with an optimal condition (60 W, 100 s) was used to ash the sample after development for removing the residual organics which have an effect on stability of the Al/$AlO_x$/Al junctions[23,24].

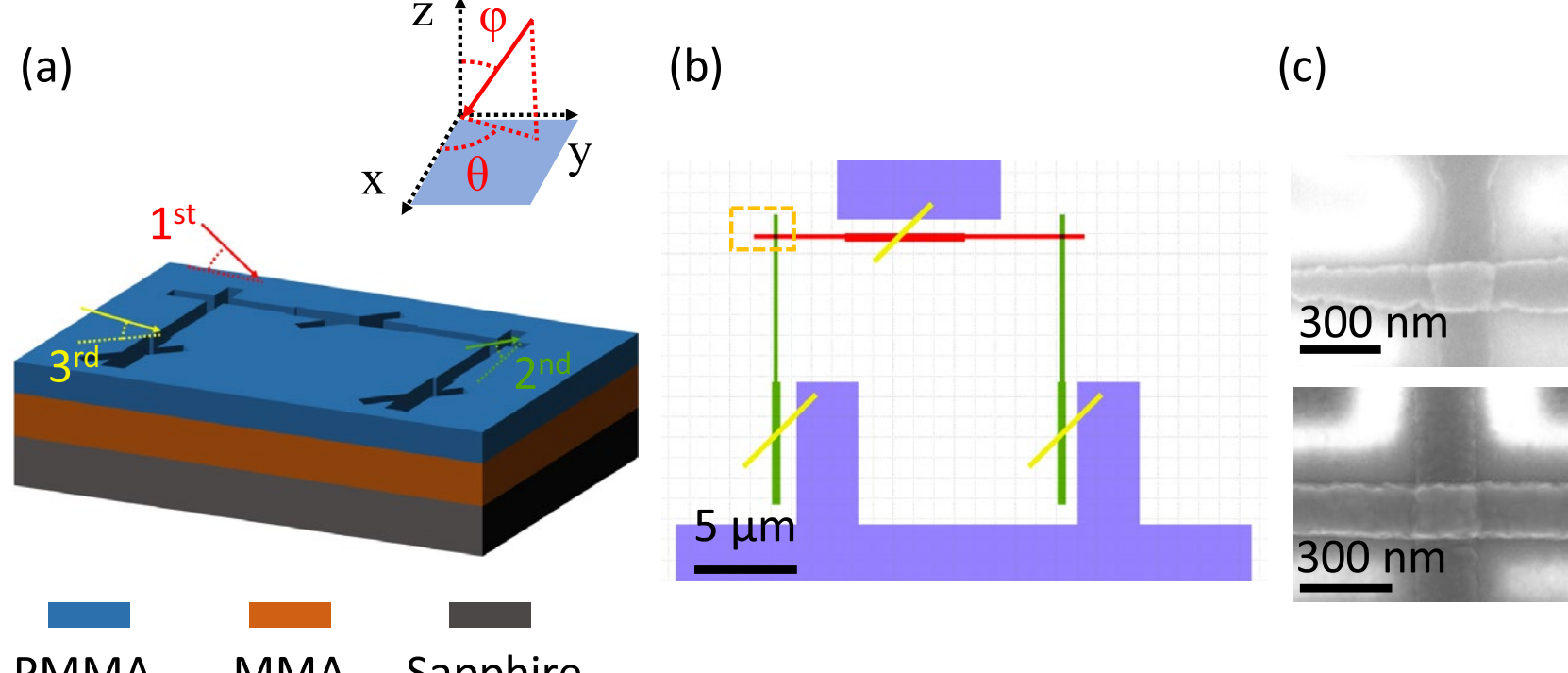

**Figure 1.** The geometry for the superconducting quantum interference device (SQUID) used in this study. (**a**) Sketch of SQUID preparation using the 'PITC' technique. The sample plane refers to the xy-plane, and the planetary and tilt angles of the sample holder are represented by θ and φ, respectively. '1$^{st}$' and '2$^{nd}$' represent the evaporation of the bottom electrode and the top electrode of the Al/$AlO_x$/Al junction, respectively. An oxidation step occurs between them, and it is not shown in this process. '3$^{rd}$' denotes the $Ar^+$ ion milling and the subsequent Al deposition used for patching. (**b**) Layout of the SQUID. The purple parts represent the test circuit structure. (**c**) Scanning electron microscope (SEM) images of an Al/$AlO_x$/Al junction. Top: The junction prepared with 10 nm conductive layer during EBL. Bottom: The junction prepared with 20 nm conductive layer during EBL.

The subsequent evaporation steps are shown in Fig. 1a, where the planetary and tilt angles (from the z-axis) of the sample holder are denoted by θ and φ, respectively. The final layout of the SQUID in this experiment is shown in Fig. 1b, and the orange rectangle marks one of the junctions. To mitigate the impact of transverse incident angle effect, which is discussed in literature[25] and leads to variation in the junctions area, the sample and sample holder were aligned under the microscope prior to introducing the sample into the ultrahigh vacuum system. After full degassing, the first Al electrode of the junctions was deposited at θ = 0° and φ = 45° to reduce the shading effects[10]. 1 nm/s deposition rate and 30 nm thickness were used. Both of the deposition angle and growth rate were optimized to achieve the best grain uniformity for the bottom electrode, which would improve the uniformity of the oxide layer in the next step[10]. After static oxidation at 5 mbar for 30 min, the second Al electrode was also deposited with 1 nm/s deposition rate at the angle of θ = 90° and φ = 45°, to the thickness of 60 nm. After removing the surface oxide layer from the sample using $Ar^+$ ion milling, aluminum was deposited at an angle of θ = 45° and φ = 60° for patching. The final step in the fabrication process is the passivation process, which involves static oxidation of these junctions at 100 mbar for 30 min. The barrier region is observed using transmission electron microscopy (TEM) (see Fig. 2), revealing a very small roughness and a steep interface between Al and O.

## Measurement and result

The critical current $I_c$ is a key parameter of Josephson junction, which depends on the junction area, oxidation condition and other chemical pollution. It is often used to characterize their quality and reliability. However, the $I_c$ have to be measured at low temperatures[26], which makes the characterization difficult. Fortunately, the $I_c$ of Josephson junction can be inferred from its normal resistance[27]. In this study, the uniformity and stability of the SQUID were characterized from their resistances, which were measured using a four-probe method to avoid the effects of contact resistance. The average resistances of these SQUIDs range from 5 kΩ to 12 kΩ.

**Uniformity.** The uniformity of the junction resistances at room temperature is shown in Fig. 3. On 15 mm × 15 mm chips, the $\sigma_{R_N}/<R_N>$ is less than 2%. On 2 inch wafers, the $\sigma_{R_N}/<R_N>$ is less than 3%. In both chip size and wafer scale, the uniformity of these junctions decreases with an increase in the junction area (Fig. 3a,d). This indicates that patterns with a larger scale exposed using low beam energy are more uniform. However, the resistance of the smallest junction size with 130 × 130 $nm^2$ still exhibits a very regular Gaussian distribution relative to the designed junction resistance, as shown in Fig. 3b,e. The spatial distribution of the junction resistances (Fig. 3c,f) shows that the relative resistance deviation is higher on the right side of the chip. This should be due to changes in the evaporation conditions as the deposition angle is changed over the wafer. The effective growth rate and shading effect can affect the grain uniformity, and the deposition angle relative to the sidewall of the resist can affect the junction area. Most of these condition variations should be improved by optimizing the evaporation procedure[24].

**Temperature dependent.** To further characterize the quality and reliability of our SQUID, we conducted resistance variation measurements with temperature. When the temperature was decreased from room temperature (300 K) to 4 K, the SQUID resistances increased by an average of 9.73%, and there was no significant change in uniformity (see Fig. 4a). The resistances almost returned to their initial values in a reversible process when the temperature rose back. The final values showed an average increase of 0.75% compared to the resistances before cooling. Figure 4b provides more detail of the resistance variation with decreasing temperature. The resistance increased rapidly from room temperature to 170 K, followed by a small increase from 170 to 4 K. Previous studies have shown that the Josephson junction resistance increases with decreasing temperature[16,28,29], which should result from the thermally fluctuation-induced tunneling conduction through hot spots in the barrier layer[28]. These hot spots should be formed in the fabrication process and result in the inhomogeneity of the oxide layer. They should also be formed by the $OH^-$ ion diffusion[30] and other chemical pollution[31,32] in the barrier layer. The fractures and other irreversible deformations result from the stress during the temperature decreasing should also lead to the increase of the junction resistance[33]. However, the variation of resistance with temperature in our SQUID is almost reversible, which is consistent with the weak insulating-like temperature dependence described by the Simmons model[34]. This confirms that the barrier layer in the junctions fabricated by our optimal process is very uniform and stable without any residual organics adsorption. It is noted that temperature depended variation of the resistance also remind us that a compensation to the junction resistance at room temperature should be needed to get accurate $I_c$ when designing the quantum devices based on the Josephson junction[16].

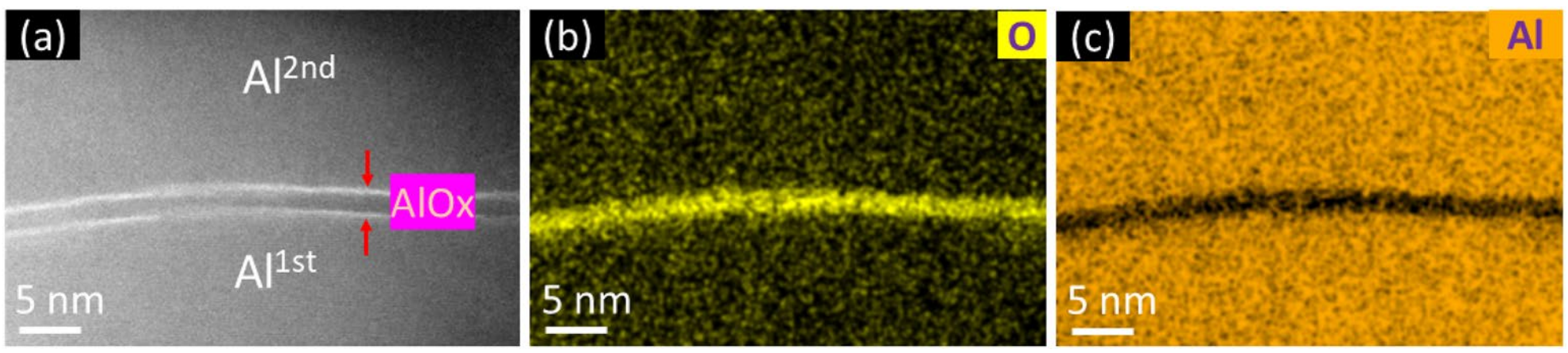

**Figure 2.** TEM images and energy dispersive x-ray spectroscopy (EDX) mapping of the Josephson junction barrier layer. (**a**) TEM images of the barrier layer. (**b**) EDX mapping of O in the barrier layer. (**c**) EDX mapping of Al in the barrier layer.

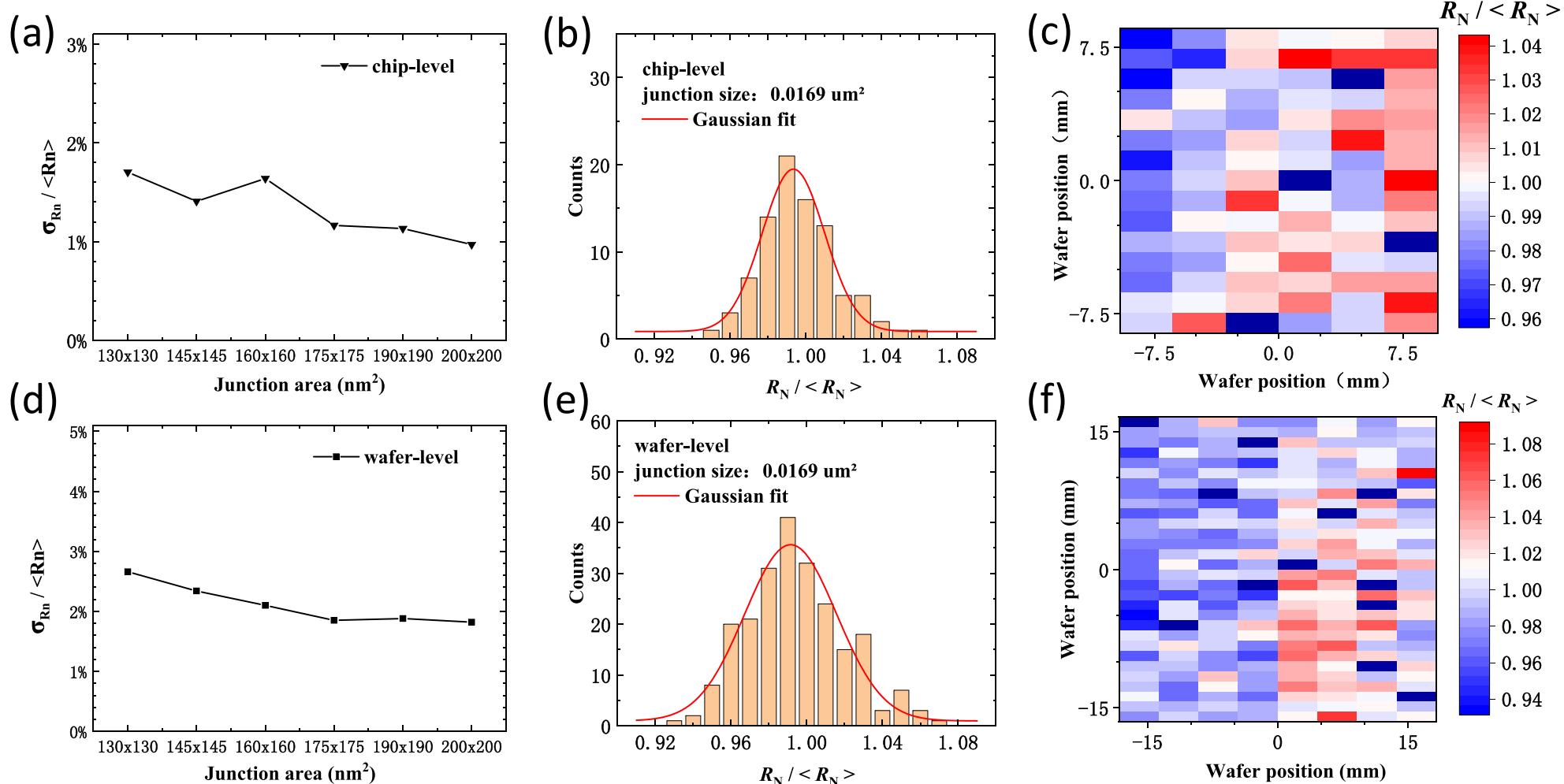

**Figure 3.** Room temperature junction resistances on 15 mm × 15 mm chips and 2 inch wafers. (**a**), (**b**), (**c**) junction resistances on the 15 mm × 15 mm chips. (**d**), (**e**), (**f**) junction resistances on the 2 inch wafers. (**a**), (**d**) $\sigma_{R_N} / < R_N >$ versus junction areas. The junction areas are 130 nm × 130 nm, 145 nm × 145 nm, 160 nm × 160 nm, 175 nm × 175 nm, 190 nm × 190 nm, 200 nm × 200 nm, and the corresponding SQUID average resistance $<R_N>$ are 11.9 kΩ, 9.63 kΩ, 7.79 kΩ, 6.53 kΩ, 5.74 kΩ, 5.09 kΩ. (**b**), (**e**) Gaussian distribution of the room temperature resistances of these junctions with junction area of 130 nm × 130 nm. (**c**), (f) Spatial distribution of the junction resistances with junction area of 130 nm × 130 nm.

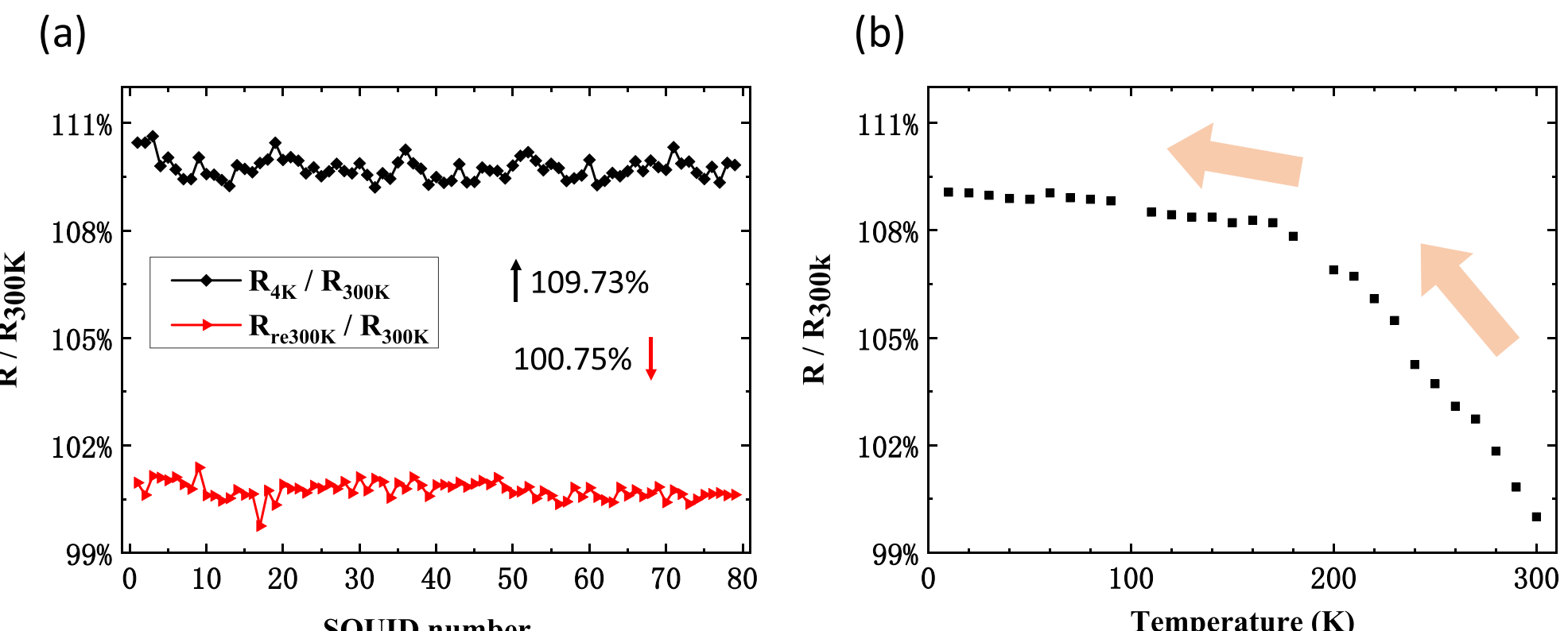

**Figure 4.** The changing of SQUID resistances response to the cooling and warming process. (**a**) The SQUID resistance variation at 4 K (black curve) and room temperature (300 K) after several cooling and warming cycles (red curve). $R_{300k}$ represents the initial SQUID resistance at room temperature. $R_{4k}$ represents the SQUID resistance at 4 K. $R_{re300k}$ represents the SQUID resistance at room temperature after one cooling and warming cycle. (**b**) The increasing process of the SQUID resistance when the temperature decreases from 300 K to 4 K. The average value of $R_{300k}$ for the SQUID is 8.92 kΩ.

**Aging.** The uniformity and stability of these junctions fabricated by optimal process are confirmed by the aging measurement furthermore. Reports have shown that the variation of oxide layer, the residual resist and other chemical pollution can cause junction parameters change[35–37]. A 15 mm × 15 mm chip with various of junctions was put into a nitrogen cabinet to simulate the aging process of superconducting qubits in a common storage environment. An average resistance variation of 1.16% relative to their initial values was observed after 100 days, as shown in Fig. 5a. There was no significant change in the uniformity of these junctions (see Fig. 5b).

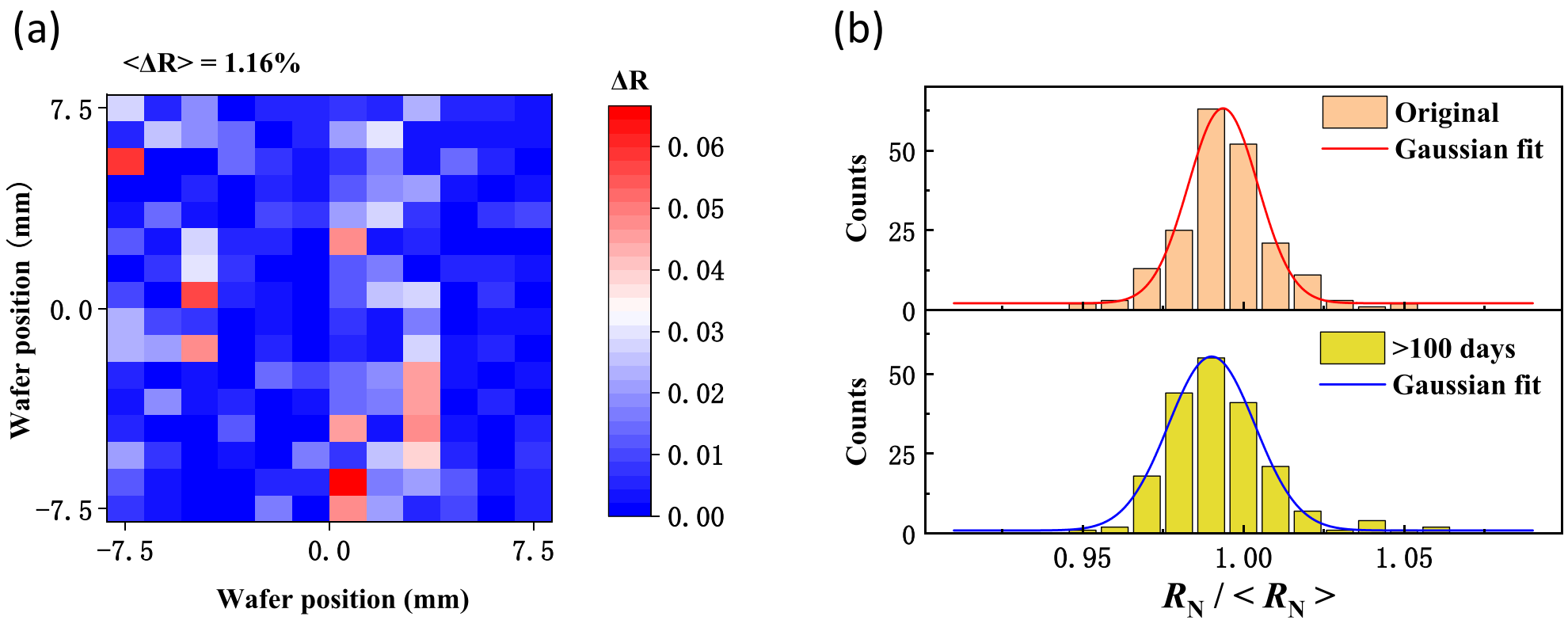

**Figure 5.** The aging of the $Al/AlO_x/Al$ junctions on sapphire substrate. (**a**) Chip map of junction resistance variation which were stored in a nitrogen cabinet for 100 days at room temperature. $\Delta R = \left|\frac{R_{N0} - R_{N100days}}{R_{N0}}\right|$, where $R_{N0}$ and $R_{N100days}$ represent the initial resistances and the resistances after 100 days, respectively. (**b**) Gaussian distribution of the initial resistances (top) and the resistances after 100 days (bottom).

## Conclusion

Motivated by fabrication of superconducting quantum processors with hundreds of qubits, a process of fabricating submicron-sized $Al/AlO_x/Al$ junctions with high uniformity and stability on a sapphire substrate was developed by using a 50 kV electron beam lithography process. These junctions with areas ranging from 0.0169 to 0.04 $\mu m^2$ exhibited $\sigma_{R_N}/ < R_N >$ values better than 1.7% on a 15 mm × 15 mm chip and better than 2.66% on a 2 inch wafer. To achieve this, a 20 nm Al layer was used as a conductive layer to reduce the charging effect during electron beam lithography. Before developing, the main Al conductive layer was removed with a TMAH dilution without attacking the photoresist, and the remaining was removed with deionized water, then the final patterns were defined, which results in sharp photoresist patterns. Then, the ashing process to remove organic residues and the Al evaporation rates related to the roughness of the bottom electrode were optimized. The junctions fabricated by this process also showed good stability. Their resistances increased at a fixed ratio of 9.73% as the temperature decreased from room temperature to 4 K, and almost returned to their initial values in a reversible process when the temperature rose back. This behavior is consistent with the Simmon model and indicates that the barrier layer of these junctions is stable and uniform. Over three months of storage in a nitrogen cabinet, these junctions had an average change in resistance of 1.16%. Our optimized process for fabricating Josephson junctions with high uniformity and stability paves the way for large-scale superconducting quantum chip fabrication on a sapphire substrate.

## Data availability

The data that support the findings of this study are available from the corresponding author upon reasonable request.

## Acknowledgements
K. L. X acknowledges support from the Youth Innovation Promotion Association of Chinese Academy of Sciences (2019319). J. G. F. acknowledges support from the Start-up foundation of Suzhou Institute of Nano-Tech and Nano-Bionics, CAS, Suzhou (Y9AAD110).

## Author contributions
Y.Z.: Investigation, Data analysis, Writing – original draft. S.L.: Investigation, Data analysis, Writing – original draft. Z.D.: Investigation, Data analysis. K.X.: Project administration, Writing – review & editing. J.F.: Project administration, Conceptualization, Supervision, Writing – review & editing. H.Y.: Supervision.

## Competing interests
The authors declare no competing interests.

## Additional information
**Correspondence** and requests for materials should be addressed to K.X., J.F. or H.Y.